\documentclass[prd,longbibliography,twocolumn,amsmath,amssymb,nofootinbib,preprintnumbers,balancelastpage]{revtex4-1}

\usepackage{feynmp-auto}
\usepackage{amsmath}
\usepackage[utf8x]{inputenc} 

\usepackage{bbm}
\usepackage{soul}	
\usepackage{xcolor}

\usepackage{epsfig}
\usepackage{color}
\usepackage{slashed}
\usepackage{comment}

\graphicspath{ {Figures/} }
\usepackage[font={small}]{caption}
\usepackage{float}
\usepackage{multirow}
\usepackage[export]{adjustbox}
\usepackage[hang, flushmargin,bottom]{footmisc} 
\usepackage{hyperref}
\hypersetup{
    pdfnewwindow=true,      
    colorlinks=true,        
    linkcolor=blue,        
    citecolor=blue,         
    filecolor=blue,         
    urlcolor=blue           
}

\usepackage{placeins}
\usepackage{enumitem}

\newcommand{\beq}{\begin{equation}}
\newcommand{\eeq}{\end{equation}}

\newcommand{\bea}{\begin{eqnarray}}
\newcommand{\eea}{\end{eqnarray}}

\newcommand{\SU}{\,{\rm SU}}

\def\beq{\begin{equation}}
\def\eeq{\end{equation}}

\usepackage[normalem]{ulem}

\DeclareUnicodeCharacter{2212}{-}

\AtBeginDocument{%
    \newwrite\bibnotes
    \def\bibnotesext{Notes.bib}
    \immediate\openout\bibnotes=\jobname\bibnotesext
    \immediate\write\bibnotes{@CONTROL{REVTEX41Control}}
    \immediate\write\bibnotes{@CONTROL{%
    apsrev41Control,author="08",editor="1",pages="1",title="0",year="1"}}
     \if@filesw
     \immediate\write\@auxout{\string\citation{apsrev41Control}}%
    \fi
}%

\usepackage{amsfonts}
\usepackage{graphicx}
\usepackage{tcolorbox}
\usepackage{amsmath}
\usepackage{slashed}
\usepackage{verbatim}
\usepackage{color}
\usepackage{array}
\newcolumntype{P}[1]{>{\centering\arraybackslash}p{#1}}
\usepackage{physics}
\usepackage[font={small, up}]{caption}
\usepackage{hyperref}	

\newcommand{\SSM}[0]{\mbox{$\Sigma\text{SM}$}}

\begin{document}

\title{ {\bf{On the $W$ mass and New Higgs Bosons}}}
\author{Pavel Fileviez P\'erez}
\email{pxf112@case.edu}
\affiliation{Physics Department and Center for Education and Research in Cosmology and Astrophysics (CERCA), 
Case Western Reserve University, Cleveland, OH 44106, USA}
\author{Hiren H. Patel}
\email{hhp25@case.edu}
\affiliation{Physics Department, Case Western Reserve University, Cleveland, OH 44106, USA}
\author{Alexis D. Plascencia}
\email{alexis.plascencia@lnf.infn.it}
\affiliation{INFN, Laboratori Nazionali di Frascati, C.P. 13, 100044 Frascati, Italy}
\begin{abstract}
We discuss the prediction of the $W$ boson mass in a simple extension of the Standard Model ($\Sigma {\rm SM}$) with a real scalar triplet.
A shift in the $W$ mass as reported by the CDF II collaboration can naturally be accommodated by the model without modifying the Standard Model value for the $Z$ mass. We discuss the main implications and the properties of the new Higgs bosons. Namely, the partial decay widths of the new charged Higgs are predicted. Furthermore, the neutral Higgs has suppressed couplings to fermions and decays predominantly into a pair of $W$ gauge bosons.
\end{abstract}
\maketitle
\hypersetup{linkcolor=blue}
%
The Standard Model (SM) of particle physics describes with high precision the properties of quarks and leptons and how they interact through the electromagnetic, weak and strong interactions. 
Nevertheless, there exist motivations for physics beyond the SM such as the existence of massive neutrinos, the dark matter and the origin of the matter-antimatter asymmetry in the Universe. 
Recently, the CDF II collaboration~\cite{CDF:2022hxs} has reported a new intriguing result for the $W$ gauge boson mass, $M_W^{\rm CDF} = 80.433 \pm 0.009 \,\, {\rm GeV}$,
which is $7 \sigma$ deviations away from the SM prediction of $M_W^{\rm SM} = 80.357 \pm 0.006 \,\, {\rm GeV}$ \cite{ParticleDataGroup:2020ssz}, that points towards a modification of the Standard Model.

The ratio between the masses of the $W$ and $Z$ gauge bosons in the Standard Model is predicted, i.e. $\cos \theta_W = M_W/M_Z$ at tree-level, because there is only a single Higgs in the theory to generate masses for the weak gauge bosons and the charged fermions. There have been different proposals to explain the value for $M_W$ reported by the CDF II collaboration using physics beyond the SM, see e.g. Refs.~\cite{Fan:2022dck,Lu:2022bgw,Athron:2022qpo,Strumia:2022qkt,Yang:2022gvz,deBlas:2022hdk,Tang:2022pxh,Cacciapaglia:2022xih,Blennow:2022yfm,Sakurai:2022hwh,Fan:2022yly,Liu:2022jdq,Cheng:2022jyi,Song:2022xts,Bagnaschi:2022whn,Bahl:2022xzi,Asadi:2022xiy,DiLuzio:2022xns,Athron:2022isz,Babu:2022pdn,Heo:2022dey,Du:2022brr,Cheung:2022zsb,Crivellin:2022fdf,Endo:2022kiw,Biekotter:2022abc,Balkin:2022glu,  Ahn:2022xeq,Han:2022juu}. {\textit{However, the simplest extension of the SM that naturally modifies the $W$ mass without affecting the $Z$ mass is the} \SSM.  This model contains a real scalar $\SU(2)_L$ triplet with zero hypercharge~\cite{Ross:1975fq,Gunion:1989ci,Lynn:1990zk,Blank:1997qa,Forshaw:2003kh,Chen:2006pb,Chankowski:2006hs,Chivukula:2007koj,Bandyopadhyay:2020otm}.

In this Letter we discuss the main properties of the physical scalars  in the \SSM~and show how the new result reported by the CDF II collaboration allows us to find the allowed value for the triplet vacuum 
expectation value in agreement with all experimental constraints. This theory predicts two extra physical scalars, a $CP$-even neutral Higgs, and a charged Higgs. 
We show that in the limit when the lightest Higgs is SM-like, the heavy $CP$-even neutral Higgs is almost fermiophobic. We discuss the numerical predictions for the Higgs decays 
and comment on the possible production channels at the Large Hadron Collider (LHC). 

The scalar sector of the \SSM~is composed of the SM Higgs,
$H \sim (\mathbf{1},\mathbf{2},1/2)$, and a real triplet, $\Sigma \sim (\mathbf{1},\mathbf{3},0)$. The relevant Lagrangian for our discussion is given by
\begin{equation}
\label{eq:lscalar}
\mathcal{L}_\mathrm{scalar} = (D_\mu H)^{\dagger} (D^\mu H) + \Tr (D_\mu \Sigma)^{\dagger}
(D^\mu \Sigma) - V(H,\Sigma), \,
\end{equation}
where $H^T=( \phi^+,\,\phi^0)$ is the SM Higgs and the real triplet is given by
\begin{equation}
\Sigma = \frac{1}{2} \left( \begin{array} {cc}
\Sigma^0  &  \sqrt{2} \Sigma^+ \\
\sqrt{2} \Sigma^-  & - \Sigma^0
\end{array} \right), \,
\end{equation}
with $\Sigma^0$ being real, $\Sigma^+=(\Sigma^-)^*$, and
$
D_\mu\Sigma =  \partial_\mu\Sigma +i g_2 [{W}_\mu,\Sigma].
$
Here $W_\mu$ and $g_2$ are the $\SU(2)_L$ gauge bosons and coupling, respectively.

The most general renormalizable scalar potential is
\begin{eqnarray}
\label{scalar1}
V(H,\Sigma) & = & - \mu^2 H^\dagger H \ + \  \lambda_0 (H^\dagger H)^2 \ - \ M^2_{\Sigma}  \Tr  \Sigma^2 \nonumber \\
& + & \lambda_1  \Tr  \Sigma^4 \ + \  \lambda_2  (\Tr  \Sigma^2 )^2 
 +  \alpha  ( H^\dagger H ) \Tr  \Sigma^2 \nonumber \\
 & + &  \beta H^\dagger \Sigma^2 H + \ a_1  H^\dagger \Sigma H .
\end{eqnarray}
Notice that in the limit $a_1 \to 0$ the scalar potential
of the theory possesses a global symmetry $O(4)_H \times O(3)_\Sigma$ and the
discrete symmetry $\Sigma \to - \Sigma$. These symmetries technically protect
the dimensionful parameter $a_1$, and the case of small $a_1$ corresponds
to a soft breaking of this symmetry.  We use the convention
$a_1>0$ by absorbing its sign into the definition of $\Sigma$.

The scalars $H$ and $\Sigma$ can have a non-zero vacuum expectation value (VEV). Consequently, we can define:
\bea
\label{eq:shifted}
H  & =&
\left(\begin{array}{c} \phi^+ \\ (v_0+h^0+iG^0) /\sqrt{2} \end{array}\right),
\eea
 and
 \bea
\Sigma  =  \frac{1}{2} \left( \begin{array}{cc}
x_0+\sigma^0  & \sqrt{2}\Sigma^+ \\
\sqrt{2}\Sigma^-  & - x_0-\sigma^0
\end{array} \right) \ ,
\eea
where $v_0$ and $x_0$ are the SM Higgs and triplet VEVs, respectively. The minimization conditions for the tree-level potential are
given by
\begin{eqnarray}
\left( - \mu^2 +  \lambda_0  v_0^2  -  \frac{a_1  x_0 }{2} +   \frac{a_2  x_0^2 }{2} \right)  v_0 &=& 0 \ ,
\label{min1}
\\
- M_{\Sigma}^2  x_0  +  b_4 x_0^3  -  \frac{a_1  v_0^2}{4 } + \frac{a_2  v_0^2  x_0}{2} & = & 0 \ ,
\label{min2}
\end{eqnarray}
and
\begin{eqnarray}
\label{eq:curve}
b_4  >  \frac{1}{8 x_0^2} \left( - \frac{a_1 v_0^2}{x_0}  +  \frac{\left(- a_1 + 2 a_2 x_0 \right)^2}{2 \lambda_0} \right). &&
\end{eqnarray}
In the above equations $a_2=\alpha + \beta/2$ and $b_4=\lambda_2 + \lambda_1/2$. For more details see the discussion in Ref.~\cite{FileviezPerez:2008bj}.

The scalars mass matrices in the basis ($h^0$ and $\sigma^0$) and ($\phi^\pm$ and $\Sigma^\pm$), are given by
\begin{equation}
\label{eq:massmtrx}
{\cal M}_{0}^2 = \left( \begin{array} {cc}
2 \lambda_0 v_0^2  &  - \displaystyle \frac{a_1 v_0}{2} + a_2  v_0  x_0 \\
-\displaystyle \frac{a_1 v_0}{2}  +  a_2  v_0  x_0  & 2 b_4  x_0^2 + \displaystyle \frac{a_1  v_0^2}{4  x_0}
\end{array} \right), 
\end{equation}
and
\begin{equation}
{\cal M}_{\pm}^2 = \left( \begin{array} {cc}
a_1 x_0  & \displaystyle \frac{a_1 v_0}{2}  \\[1ex]
\displaystyle \frac{a_1 v_0}{2} & \displaystyle \frac{a_1 v_0^2}{4 x_0}
\end{array} \right) \ .
\end{equation}

In terms of the electroweak gauge eigenstates, the physical mass eigenstates are given by
\begin{eqnarray}
\label{eq:neutralmix}
\left( \begin{array}{c} h \\ H \end{array} \right) & = & \left(\begin{array}{ccc} \cos \theta_0 & \sin \theta_0 \\ - \sin \theta_0 & \cos \theta_0\end{array} \right) \left( \begin{array}{c} h^0 \\ \sigma^0 \end{array} \right) \,,\\
& & \nonumber \\
\label{eq:chargemix}
\left( \begin{array}{c} H^\pm \\ G^\pm \end{array} \right) & = & \left(\begin{array}{cc} -\sin \theta_+ & \cos \theta_+  \\ \cos\theta_+ & \sin \theta_+  \end{array} \right) \left( \begin{array}{c} \phi^\pm \\ \Sigma^\pm \end{array} \right) \ ,
\end{eqnarray}
where the mixing angles are
\begin{equation}
\label{eq:thetazero}
\tan 2 \theta_0 =  \frac{4 v_0 x_0(-a_1 + 2 x_0 a_2)}{8 \lambda_0 v_0^2 x_0 - 8
b_4 x_0^3 - a_1 v_0^2},
\end{equation}
\begin{equation}
\tan 2 \theta_+ =  \frac{4 v_0 x_0}{4 x_0^2-v_0^2}\,.
\label{theta+}
\end{equation}
The dimensionless couplings $\lambda_0, a_2,$ and $b_4$ need to remain perturbative, and hence, from Eq.~\eqref{eq:thetazero} we see that in the limit $v_0 \gg x_0$ we obtain $\theta_0 \ll 1$. In this limit, the mass for the physical scalars $h$, $H$, and $H^\pm$ masses are given by
\bea
\label{mneut1}
M_{h}^2 & = & 2 \lambda_0 v_0^2,  \\
M_{H}^2 & = & 2 b_4 x_0^2 + \frac{a_1 v_0^2}{4 x_0}, \\
\label{eq:mpls1}
M_{H^{\pm}}^2 &=&  a_1 x_0 \left( 1 + \frac{ v_0^2}{4 x_0^2} \right),
\eea
and $h$ becomes SM-like. Notice that since $v_0 \gg x_0$ one finds $M_{H^\pm} \approx M_H$. This relation has important implications when we study the Higgs decays. 


The recently reported value by the CDF II collaboration of the $W$ boson mass is~\cite{CDF:2022hxs}
\beq
M_W^{\rm CDF} = 80.433 \pm 0.009 \,\, {\rm GeV},
\eeq
which is $7 \sigma$ deviations away from the SM prediction of
\beq
M_W^{\rm SM} = 80.357 \pm 0.006 \,\, {\rm GeV},
\eeq
that comes from an electroweak fit to different observables~\cite{ParticleDataGroup:2020ssz}. 
We mention that the CDF II measurement has a roughly $3.6\sigma$ tension with previous measurements of 
$M_W$ performed by ATLAS~\cite{ATLAS:2017rzl} and LHCb~\cite{LHCb:2021bjt}. 
However, the CDF II measurement has a much smaller uncertainty.

\begin{figure}[t]
\centering
\includegraphics[width=0.9\linewidth]{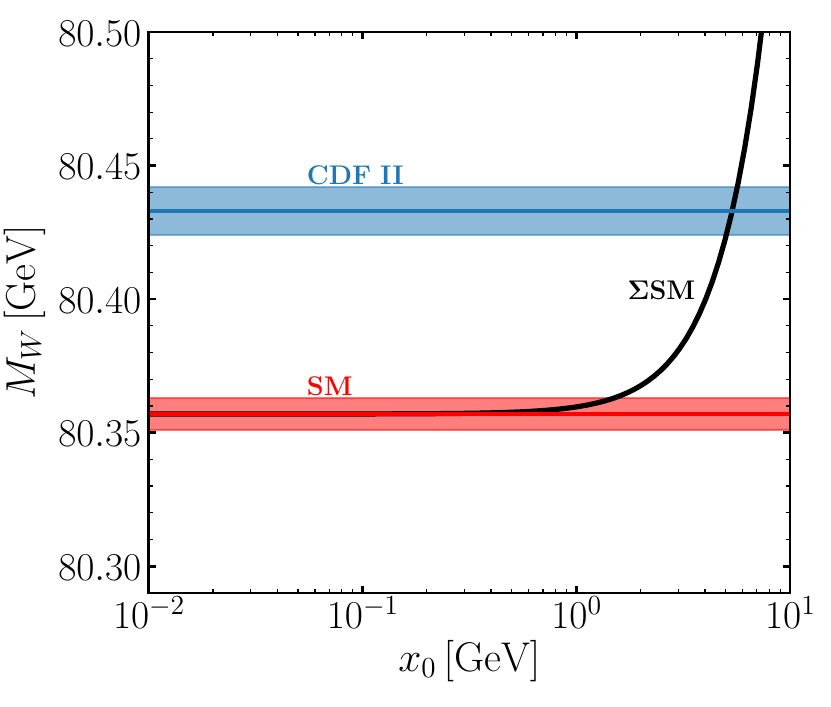}
\caption{Correlation between the $W$ boson mass, $M_W$, and the VEV of $\Sigma$, $x_0$, shown by the black line. The region shaded in red corresponds to the SM prediction and the region shaded in blue is the measurement reported by the CDF II collaboration.}
\label{fig:MWx0}
\end{figure}

In this model the $W$ mass is given by
\beq
M_W^2 = \left(M_W^\text{SM}\right)^2 + g_2^2 x_0^2,
\eeq
where the first term is the SM value and the second term is the tree-level contribution coming from the triplet.

In this theory, the vacuum expectation value of $\Sigma$ does not contribute to the $Z$ boson mass since there is no coupling of the form $(\sigma^0)^2 Z_\mu Z^\mu$, while there is an interaction $(\sigma^0)^2 W^-_\mu W^{+\mu}$. Therefore, we assume the SM value for the $Z$ boson mass. In order to address the deviation in $M_W$ and using  \mbox{$g_2(\mu\!=\!M_Z)\!=\!0.65171$}~\cite{ParticleDataGroup:2020ssz}, we find that at tree-level 
\beq
x_0 = 5.4 \,\, {\rm GeV}.
\label{eq:x0}
\eeq
Then, Eq.\eqref{theta+} immediately implies a prediction of the mixing angle in the charged sector
\begin{align}
\theta_+ & = -0.044.
\end{align}
In Fig.~\ref{fig:MWx0} we show the correlation between the $W$ boson mass and the VEV of the triplet $x_0$. The CDF II measurement and SM prediction are indicated by a blue and red bands, respectively.

{The mixing angle in the neutral sector is not predicted; however, we can use the following relation
\beq
\cos{\theta_0} \sin{\theta_0}  \left( M_H^2 - M_h^2 \right) = v \left(  a_2 x - \frac{a_1}{2} \right),
\eeq
to find the allowed values for $\sin\theta_0$. In Fig.~\ref{fig:theta0} we show the neutral mixing angle as a function of the quartic coupling $a_2$, the latter is shown in the perturbative range $\lesssim  2\sqrt{\pi}$. The typical values for this mixing angle are $\theta_0 \simeq \mathcal{O}(10^{-2})$. The lines with different colors correspond to different masses $M_{H^+} \simeq M_H$. In the limit $M_H \gg M_h$ then $\theta_0 \to \theta_+$ since the $a_1$ term dominates in Eq.~\eqref{eq:thetazero}. }

\begin{figure}[t]
\centering
\includegraphics[width=0.9\linewidth]{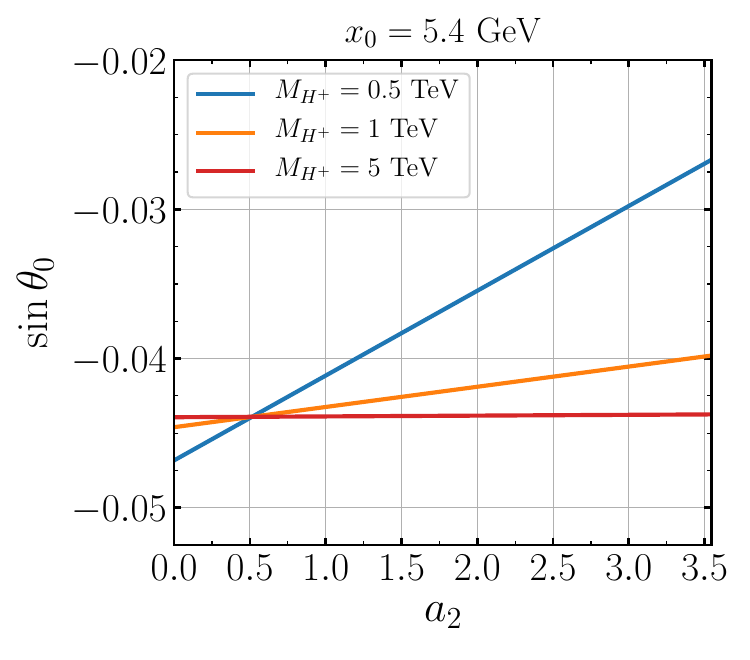}
\caption{Mixing angle for the neutral Higgs bosons as a function of the quartic coupling $a_2$. Different colors correspond to different values of $M_{H^+} \simeq M_H$.}
\label{fig:theta0}
\end{figure}
%
This model predicts a new $CP$-even Higgs, $H$, and a charged scalar, $H^{\pm}$. 
The Feynman rules involving the new scalars in the $\theta_0 \to 0$ limit are presented in Table~\ref{tab:feynman}. In this case, $H$ does not couple to SM fermions nor to the $Z$ gauge boson. 
Consequently, the dominant two-body decay for the new $CP$-even Higgs is $H \to WW$ since $H$ becomes almost fermiophobic. On the other hand, the main decay channels of the charged Higgs are
$H^+ \to \tau^+ {\nu}_\tau$, $H^+ \to t \bar{b}$, $H^+ \to W^+ Z$, and $H^+ \to h W^+$.

As we have discussed in the previous section, the CDF II value for $M_W$ fixes values for $x_0$ and $\theta_+$. Therefore, the couplings to $H^+$ are determined and we can predict its branching ratios. In Fig.~\ref{fig:chargedBR} we present the branching ratios as a function of $M_{H^+}$ for the parameters that are consistent with the CDF II measurement. Above $M_{H^+}>600$ GeV, the modes $H^+ \to h W^+$ and $H^+ \to Z W^+$ dominate the decay, and asymptotically approach each other on account of the Goldstone equivalence theorem.

{One of the main production channels at hadron colliders for the neutral Higgs is $u \bar{d} \to W^* \to W^+ H$ and it can also be produced through vector boson fusion. However, as we can see from Table~\ref{tab:feynman} the $WWH$ vertex is suppressed by $x_0$. Consequently, there are better prospects to observe the production of the neutral and charged Higgs bosons $u \bar{d} \to W^*\to H^+H$. The charged Higgs can be pair produced through the Drell-Yan process $q \bar{q} \to \gamma, Z  \to H^\pm H^\mp$ that gives a cross-section of $\mathcal{O}(1)$ fb at the LHC with $\sqrt{s}=13$ TeV for $M_{H^+} \simeq 500$ GeV. The collider phenomenology has been extensively studied in Refs.~\cite{FileviezPerez:2008bj,Chiang:2020rcv}.

\begin{figure}[t]
\centering
\includegraphics[width=0.9\linewidth]{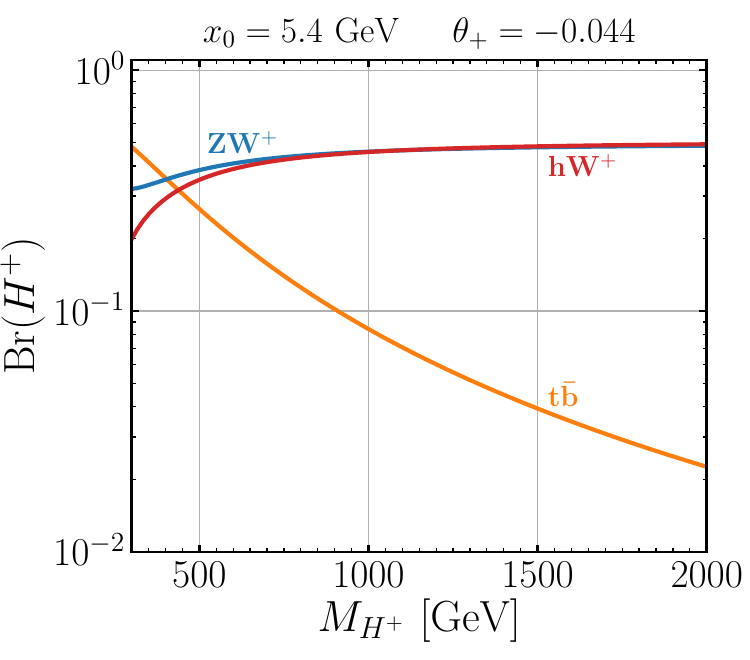}
\caption{Prediction of the branching ratios for the charged Higgs as function of $M_{H^+}$, we fix $x_0 = 5.4\text{ GeV}$.}
\label{fig:chargedBR}
\end{figure}

The charged Higgs can be single produced through the process $g b \to H^- t$, this channel with the subsequent decay $H^-\!\to \bar{t} b$ is been searched for at the LHC~\cite{Chen:2022zsh}, in our scenario this decay channel is relevant for $M_{H^+}\lesssim 500$ GeV. Furthermore, the charged Higgs can be produced through vector boson fusion and there are ongoing searches at the LHC for the process  $p p \to H^\pm jj \to W^\pm Z jj$~\cite{CMS:2018ysc,CMS:2021wlt,ATLAS:2022jho}; however, the $ZW^\pm H^\mp$ vertex is either suppressed by $x_0$ or $\sin \theta_+$. Therefore, there are better prospects to observe pair-production through vector boson fusion $pp \to H^+ H^- jj$ since the quartic coupling $W^+W^-H^+H^-$ is not suppressed. Regarding the mass of the new Higgs bosons, there exists an upper bound that comes from the perturbative unitarity of the $WW$ scattering cross-section~\cite{Chivukula:2007koj}. For the value of $x_0$ required to explain the $W$ boson mass, we obtain $M_{H,H^+}<40$ TeV.}

In this Letter, we showed that the simplest extension of the SM model that automatically explains the CDF II result for $M_W$ is the \SSM.
Consistency with CDF II result in this model fixes the branching ratios of the charged Higgs, and are shown in Fig.~\ref{fig:chargedBR}.
We also noted that the heavy $CP$-even Higgs is fermiophobic-like, and the theory predicts that its dominant decay channel is $H\rightarrow WW$. We briefly discussed the main production channels at the LHC.    

\vspace{1.5cm}
{\textit{Acknowledgments:}}
{\small{A.D.P. is supported by the INFN “Iniziativa Specifica” Theoretical Astroparticle Physics (TAsP-LNF) and by the Frascati National Laboratories (LNF) through a Cabibbo Fellowship, call 2020. Futhermore, A.D.P. acknowledges helpful discussions with Enrico Nardi.}}


\label{sec:appFR}
\begin{center}
\begin{table}[h]
\begin{tabular}[t]{|c||c|c|}
\hline Interaction & Feynman Rule \\
\hline
$h f\bar{f}$&$i(M_f/v_0)$\\[1ex]
$H^+ \bar{\nu}_i e_i$ & $-i \frac{\sqrt{2}}{v_0} M_e^i \sin \theta_{+}  P_R  $ \\[1ex]
$H^+ \bar{u} d$ & $-i \frac{\sqrt{2}}{v_0} \sin \theta_{+}  \left( - M_u  V_{\rm CKM} P_L +  V_{\rm CKM}  M_d  P_R \right)$ \\[1ex]
\hline \hline
\hline 
$ZZ h$&$(2iM_Z^2/v_0) g^{\mu\nu}$\\[1ex]
$ZW^\pm H^\mp$&$ig_2\big(-g_2x_0c_+c_w+\frac{1}{2}g_Y v_0s_+s_w\big)g^{\mu
\nu}$\\[1ex]
$W^+W^- h$&$ig^2_2\big(\frac{1}{2}v_0 \big)g^{\mu\nu}$\\[1ex]
$W^+W^-H$&$ig^2_2\big(2x_0\big)g^{\mu\nu}$\\[1ex]
$\gamma H^+H^-$&$ie\,\big(p'-p\big)^\mu$\\[1ex]
$ZH^+H^-$&$i\big(g_2c_w-\frac{M_Z}{v_0}s_+^2)\big(p'-p\big)^\mu$\\[1ex]
$W^\pm h H^\mp$&$\pm ig_2\big(\frac{1}{2}s_+ \big)\big(p'-p\big)^\mu$\\[1ex]
$W^\pm H H^\mp$&$\pm ig_2 c_+ \big(p'-p
\big)^\mu$ \\[1ex]
\hline
\end{tabular}
\caption{Feynman Rules in the limit when $h$ is SM-like ($\theta_0 \to 0$)}
\label{tab:feynman}
\end{table}
\end{center}


\bibliography{Triplet}

\end{document}